\documentstyle[12pt]{article}
\begin{document}
\vspace*{1cm}
{\hfill \Large\bf Photonic production of {\it\bf $B_c$}-mesons. \hfill }
\vspace*{1cm}
\begin{center}
A.V.Berezhnoy\footnote{Moscow Institute of Physics and Technology},
A.K.Likhoded\footnote{E-mail:LIKHODED@MX.IHEP.SU}, M.V.Shevlyagin
\end{center}
\begin{center}
Institute for High Energy Physics, Protvino, 142284, Russia
\end{center}
\vspace*{1cm}
\begin{abstract}
The cross section of the $\gamma \gamma \rightarrow B_c(B_c^*) \bar b  c $
process is calculated. It is shown that near threshold the pseudoscalar state
production is much suppressed with respect to the vector one. At large
energies their ratio becaims $\sigma(B_c^*)/sigma(B_c)\sim 4$.
The process of heavy quark recombination dominates in the production of
$B_c(B_c^*)$ states. The fragmentation process $b\to B_c$ comes to play at high
$p_t$ values only, while its contribution will remain nondominant.
\end{abstract}
\vspace*{0.2cm}
\noindent
\newpage
\section*{Introduction}
The study of the heavy quarkonium properties plays an important role in a
verification of ideas about strong interactions and an investigation of
production mechanisms of these particles allows one to test QCD predictions
in the form of the perturbative expansion. So, the mesons with the mixed
flavours  ($B_c(B_c^*)$) take a particular place due to both a specific
character of their production, being the result of production of two
pairs of the heavy quarks, and properties of the particle decays, where
hadronic channels with the heavy quark annihilation are excluded.

At the present, the mechanism of the $e^+e^- \rightarrow B_c(B_c^*)+X$
production is studied in details and some initial results related to
the hadronic
production of these particles are obtained [1-5]. The low value of the
$B_c$-meson yield ($\sim 10^{-3}$ of the  $b\bar b$-pairs production
cross section) strongly decreases the possibilities for its experimental
research.
So, the expected number of the $B_c$-mesons in the $Z_0$-boson pole is
about $10^2$ per $10^6$ $Z^0$-boson events. A real number of the observed
events with account of the particular decay modes will be less than
the expected one. Therefore the perspectives for the studies of the
$B_c$-meson properties at $e^+e^-$-colliders are very restricted.

In the present paper we would like to pay an attention to an another
possibility. At present, the future $\gamma \gamma$-colliders with the high
luminosity ($\sim 10^{34}$ $á¬^{-2}ᥪ^{-1}$) are intensively discussed.
The calculated cross section of the $B_c$ production at the moderate
energies $\sqrt{s}=20$ GeV is about 1~pb and, in experiments at the
$\gamma \gamma$-colliders, allows one to estimate the $B_c$ yield as
$\sim 10^5$ per year, so that this level is approximately two orders of
magnitude greater than the yield in the experiments at LEP. From the
other hand, the $B_c$ production in the $\gamma \gamma$-collisions has an
independent interest in the view of the theoretical understanding
of the $B_c$ production in the perturbative QCD, since in the description
complexity this reaction takes an intermediate place between  $e^+e^-$
production and hadronic one.

\section{$B_c$-meson cross section}
Calculating the production of the $S$-wave states of $B_c$-mesons
we assume that the  production process consists of  two subprocesses:
the hard production of the four free quarks with the particular kinematics and
the soft fusion of the $b$- and $\bar c$-quarks into the $B_c$-meson.

Calculating the matrix element corresponding to the first subprocess,
we have assumed that its characteristic virtualities are large, and, hence,
one can use the methods of the perturbative QCD. The part, corresponding to
the latter subprocess of soft fusion, results in the $(b\bar c)$
quarkonium wave function at the origin $\Psi (0)$, calculated in a
nonrelativistic potential model.

As it was shown in papers  [1,5], in such approach the matrix element of the
$B_c$-meson production is expressed through the matrix elements of the free
quarks as:
\begin{equation}
\label{M}
M(\lambda_i)=\frac{\sqrt {2M}}{\sqrt {2m_b} \sqrt {2m_c}} \Psi (0)
\sum_{h,\bar h,q, \bar q }^{}P_{h,\bar h}M_{h,\bar h,q, \bar q}(\lambda_i)
\frac{\delta_{q, \bar q}}{\sqrt{3}}
\end{equation}
with the following relation between the momenta
($p_b$, $p_{\bar c}$) of the
quarks composing the $B_c$-meson  and the momentum of $B_c$-meson

$$ p_b =\frac{m_b}{M}P, \qquad p_{\bar c} =\frac{m_c}{M}P,$$

where $m_b$ is the $b$-quark mass, $m_c$ is the $c$-quark mass and
$M=m_b+m_c$ is the $B_c$-meson mass.

The summin in (\ref{M}) runs over the helicities ($h,\bar h$)
and colour ($q,\bar q$) indices of the quark and the antiquark, that
compose the $B_c$-meson. The helicities of remaining fermions are
symbolically denoted as $\lambda_i$. The diagrams, corresponding to the
amplitudes $M_{h,\bar h,q, \bar q}(\lambda_i)$ are shown in fig.1.

The projection operators $P_{h,\bar h}$ determining the prescribed value of
the $B_c$-meson spin, have the following form [1,5]:
$$ P_{h,\bar h}=\frac{1}{\sqrt 2}{(-1)}^{ {\bar h}-1/2 }\delta_{h,\bar h} $$
--- for $S_0$-state

$$ P_{h,\bar h}=|h-\bar h|+\frac{1}{\sqrt 2}\delta_{h,\bar h} $$
--- for $S_1$-state.

The value of the wave function at the origin $\Psi (0)$ is calculated
in a nonrelativistic potential model as well as in the QCD sum rules [11]
and it is related with the decay constant $f_{B_c}$ of the pseudoscalar
$B_c(0^{-})$-meson and the constant $f_{B^*_c}$ of the vector
$B^*_c(1^{-})$-meson in the following way:

$$ \Psi (0)=\sqrt{\frac{M}{12}}f_{B_c},$$ £¤¥
$$f_{B_c}=f_{B^*_c}=570{\rm MeV}$$

Different potentials give  close values for the masses of the
lowest $B_c$-states. So, the pseudoscalar $0^{-}(1S$-state) mass is
evaluated as $M=6.3$ GeV. In this connection, in the calculation of the
cross section for the production of the $1S$-levels bound states of the
$B_c$-mesons, the values of the $b-$ and $c-$quark masses are taken slightly
greater than the free quark masses in the $b{\bar b}c{\bar c}$ production,
so has $m_b=4.8$ Ē' and $m_c=1.5$ GeV.

The calculation procedure used here is based on the direct numerical
evaluation of the amplitudes with its subsequent squaring and the
Monte-Carlo integration over the phase space. In this processes the
the colour factor is trivial and its calculation results in the multiplication
of the amplitude squared by the factor 16/9.

The calculations are made in two independent ways so the Weyl representation
is used in the first one, and the Dirac representation is used in the second
way. Both programs have been tested on the invariance with respect to
the Lorentc boosts along the beam axis, the spatial rotations around the
beam axis and the permutations of the initial photons. Both methods give
the same result.

Supposing $\alpha_s =0.2 $, $\alpha=1/128$, we present the
$B_c(B_c^*)$ production cross sections in table \ref{sig}.

çâ® $\alpha_s =0.2 $, $\alpha=1/128$.

\begin{table}[t]
\caption{Photonic $B_c(B_c^*)$ production cross section. (Bracketed is
the one standard deviation error on the last digit.) }
\label{sig}
\begin{center}
\begin{tabular}{|c|c|c|c|c|}    \hline
 $                  $ & 15 GeV & 20 GeV & 40 GeV & 100 GeV \\ \hline
 $\sigma_{B_c}, pb$&$5.14(1)\cdot 10^{-3}$&  $3.82(1)\cdot 10^{-2}$&
 $6.72(3)\cdot 10^{-2}$ &  $2.45(3)\cdot10^{-2} $   \\   \hline
 $ \sigma_{B_c^*}, pb$&$ 2.84(1)\cdot 10^{-1}    $
&$5.98(2)\cdot 10^{-1}$& $3.97(2) \cdot 10^{-1}$& $1.07(3)\cdot 10^{-1}$\\
  \hline
\end{tabular}
\end{center}
\end{table}

This table and fig.2 contain the cross sections
$B_c$ and $B_c^*$ production in comparison with
the cross section of the photonic production of the $b\bar b$-pair.
One can see that near threshould the pseudoscalar state production is
suppressed
in comparison with the production of the vector one, so at $\sqrt{s}=15$ GeV
one has $\sigma_{B_c^*}/ \sigma_{B_c} \sim 55$.
Such behavior of the
$\sigma_{B_c^*}/ \sigma_{B_c}$ ratio has been noted in [6], where the
strong suppression of the pseudoscalar meson pair production with respect
to the vector one takes place in the quark-antiquark annihilation.
At large energies of the initial photons this ratio decreases and becomes
$\sigma_{B_c^*}/ \sigma_{B_c} \sim 4$. As one can see from fig.2, the
inclusive cross sections $\sigma_{B_c}$ and $\sigma_{B_c^*}$ have the
maximum at $\sqrt{s}=20-30$~GeV and with the $s$ growth they
fall like the total cross section for the heavy quarks $\sigma_{b \bar b}$
production.

The distributions
$\frac{1}{\sigma_{B_c}} \cdot \frac{d\sigma_{B_c}}{dz}$ and
$\frac{1}{\sigma_{B_c^*}} \cdot \frac{d\sigma_{B_c^*}}{dz}$ over
the variable $z=\frac{2|\vec p|}{\sqrt{s}}$, with $\vec p$ being
the meson momentum, are shown in figs.3.4 for three values of the total
energy. As follows from these figures the scaling in these distributions is
broken: with the energy growth the shift onto the low $z$ values takes
place. Note, the analogous picture has been observed in the gluonic
production of $B_c$-mesons [8].

The figs.5a,b, where for the energies are  20 and 100 GeV the distributions
$\frac{1}{\sigma} \cdot \frac{d\sigma}{d\cos \Theta}$ are presented for
$\Theta$, being the angle between the meson and the beam axis, show that
both the $B_c$ and $B_c^*$ production near the threshould are practically
isotropic and at the large energies the mesons are produced in the
directions close to the directions of the initial photons.

\section{Fragmentation contributions into the $B_c$-meson production cross
section}

Note that the detailed consideration shows that in the matrix element of the
$\gamma \gamma \rightarrow b \bar b c \bar c$ process and hence in the
$\gamma \gamma \rightarrow B_c \bar b  c$ matrix element one can distinguish
three groups of contributions which are separately gauge invariant under
both the gluon field transformation and the photon one. The first group
of contributions is composed of the diagrams when the quark production
is independent (we will label these diagrams as the recombination diargams),
the second group consist of the diagrams where the $c \bar c$ pair is
produced from the $b$-quark line (we will mark these diagrams as the
$b$-quark fragmentation diagrams, their contribution will be denoted as
$\sigma^{b-frag}$), the the third group contains the diagrams with the
$b\bar b$ pair
production from the $c$-quark line, so that they are $c$-fragmentation
diagrams with the corresponding contributions denoted as $\sigma^{c-frag}$.

In refs. [4,9] the assumption was offered that the $b$-fragmentation
contribution has to dominate at large values of the $B_c$ transverse
momentum, independently of the type of the process.
So the approximate equation has to be valid:
\begin{equation}
\label{pt}
\frac{d\sigma^{q-frag}_{B_c}}{dP_t}=
\int \limits_{\frac{2P_t}{\sqrt{s}}}^{1}
\frac{d\sigma_{q \bar q}}{d k_t}(\frac{P_t}{z}) \cdot
\frac{D_{q\rightarrow B_c}(z)}{z}dz,
\end{equation}
where $\frac{\sigma_{q \bar q}}{dk_t}$ -- is the differential cross section
for the production of the fragmenting $q$-quark in the Born approximation,
$k_t$ is its transversal momentum, and
$D_{q\rightarrow B_c}(z)$ is the function of the $q\rightarrow B_c+X$
fragmentation.

Remind that in the $e^+e^-$-annihilation the $b$-quark fragmentation dominates
and the $c$-quark fragmentation contribution is suppressed by two orders of
magnitude. In the $\gamma \gamma$-interactions, the $c$-quark fragmentation
contribution is enlarged due to the quark charge ratio $(Q_c/Q_b)^4=16$
and, therefore we can not neglect it (as one does in $e^+e^-$-annihilation).
 Note further, that the $c$-quark fragmentation
contribution and the $b$-quark fragmentation one are related to each other
by the simple permutation of the quark masses and charges
($m_c\leftrightarrow m_b$ and $Q_c\leftrightarrow Q_b$) (\ref{pt}).

In the exact calculation of the fragmentation contribution, the matrix
elements are obtained by the same substitutions, while the results of the
$b$- and $c$-fragmentations are shown on fig.6 as the dotted and dashed
histograms, respectively. This test has been made for the exclusion
of error in the calculation. Comparing the dashed line with curve 2 on fig.6
we are convinced that the real contribution of the $c$-fragmentation
diagrams is much greater than the expected value, presented by eq.(\ref{pt}),
and, therefore, this contribution into the cross section can not be explained
in the framework of the fragmentation mechanism. Moreover, in contrast to
this mechanism, the $c$-fragmentation contribution is greater than the
$b$-fragmentation one at any transverse momentum, so the former comes close
to the latter only at $P_t \sim 40$ GeV. Note that in the derivation of the
fragmentation function from (\ref{pt}), one does not take into account the
contribution of the diagrams where the gluon vertex stands between the
propagators of the fragmenting quark (the middle diagram on fig.1).
For such diagrams of the $B_c$ production, the kinematical situation is
possible when one of the fermion propagators is close to the mass shell,
so that
this configuration results in the large contribution of the $c$-fragmentation.

The analogous situation is valid for the vector state, as fig.7 exhibits.

\section*{Conclusion}
We have calculated the two-photonic cross section for the $B_c$ production.
In its maximum at the energy 20-30 GeV the total cross section, including
the $B_c^*$ and corresponding antiparticle production, is about 1 pb.
This correspond to $10^5$ $B_c$, produced at the $\gamma \gamma$-collider
with luminocity of $10^{34} cm^{-2}sec^{-1}$.
At large energies, the cross section
falls like the $b\bar b$-pair production one. The $B_c$ production
mechanism is close to that of in the gluon-gluon interactions [8],
and it does not come to the simple $b$-quark fragmentation. The latter
is not large part of the cross section even at the large transverse
momenta.

We thank V.V.Kiselev and O.P.Yushchenko for useful discussions and valuable
support in our work.

The work of M.V.Shevlyagin was supported by the Russian
Fund of Fundamental Researches (project number 93-02-14456).

\newpage
\section*{\ \ \ \ ‹¨â¥à âãà }
\begin{enumerate}
\item[1.]
{\it Ji C.R., Amiri F.}//Phys.Rev. 1987. V. D35. P. 3318.,
Phys.Lett. 1987. V. B195. P. 593.
\item[2.]
{\it Chang C.-H. and  Chen Y.-Q.}//Phys.Rev. 1992. V. D46. P. 3845.
\item[3.]
{\it Kiselev V.V., Likhoded A.K. and Shevlyagin M.V.}//Preprint IHEP 93-58.
 Protvino, 1993; //Sov. Jad. Phys. 1994. v. 57 p. 103.
\item[4.]
{\it Braaten E., Cheung K., Yuan T.C.}//Phys.Rev. 1993. V. D48 P. 4230;
Phys.Rev. 1993. V. D48. P. 5049.
\item[5.]
{\it Kiselev V.V., Likhoded A.K., Shevlyagin M.V.}//Preprint
IHEP 94-10. Protvino, 1994.
\item[6.]
{\it Kiselev V.V., LIkhoded A.K. and Tkabladze A.B.}//Sov. Jad. Phys. 1987. v.
46. p. 934.
\item[7.]
{\it Chen. Y.-Q.}//Phys.Rev. 1993. V. D48. P. 5181.
\item[8.]
{\it Berezhnoy A.V., Likhoded A.K., Shevlyagin M.V.}//Preprint
IHEP 94-48. Protvino, 1994.
\item[9.]
{\it Cheung K., Yuan T.C.}//Phys.Lett. 1994. V. B325. P. 481.
\item[10.]
{\it Slabospitsky S.R.}//Preprint IHEP 94-53. Protvino, 1994.
\item[11.]
{\it Kiselev V.V.}//Nucl.Phys. 1993. V. B406. P. 340.
\end{enumerate}

\newpage
\section*{Figure captions}
\begin{itemize}
\item[¨á. 1]
The diagram type in the photonic production of the four free quarks
(Solid line is the quark, waved one is the photon, spiral line is the gluon).
\item[¨á. 2]
The total cross section (in pb) for the process of
$\gamma \gamma \rightarrow B_c \bar b c$ (empty triangle)
and $\gamma \gamma \rightarrow B_c^* \bar b c$
(solid triangle) versus the photon energy. For the comparison, the dependence
of the $b \bar b$-pair production cross section (solid line) is shown.
\item[¨á. 3]
The $z$-distribution, normalized by unit, for the
$B_c$-meson production at the energies 20 GeV (dotted line),
40 GeV (dashed line) and  100 GeV (solid line).
\item[¨á. 4]
The same as in Fig.3, but for $B_c^*$ production.
\item[¨á. 5]
The normalized by unit distribution over $\cos \Theta$ for the photonic
production at 20 GeV (dashed line) and 100 GeV (solid line) for:
a) $B_c$; b) $B_c^*$.
\item[¨á. 6]
The gauge invariant contributions in $B_c$-meson production at 100 GeV as
a function of transverse momentum:
all diagrams (nonlabelled solid line), $b$-fragmentation diagrams
(dashed line), $c$-fragmentation diagrams (dotted line).
Line 1 corresponds to the prediction of the $b$-quark fragmentation mechanism,
and line 2 is one for the $c$-quark.
\item[¨á. 7]
The same as on Fig.6, but for $B_c^*$.
\end{itemize}

\end{document}